\documentstyle[aaspp4,flushrt]{article}
\input psfig.sty

\makeatletter

\@ifundefined{chapter}{\def\thebibliography#1{\section*{References\@mkboth
  {REFERENCES}{REFERENCES}}\list
  {\relax}{\setlength{\labelsep}{0em}
        \setlength{\itemindent}{-\bibhang}
        \setlength{\itemsep}{0pt}
        \setlength{\parsep}{0pt}
        \setlength{\leftmargin}{\bibhang}}
    \def\newblock{\hskip .11em plus .33em minus .07em}
    \sloppy\clubpenalty4000\widowpenalty4000
    \sfcode`\.=1000\relax}}%
{\def\thebibliography#1{\chapter*{Bibliography\@mkboth
  {BIBLIOGRAPHY}{BIBLIOGRAPHY}}\list
  {\relax}{\setlength{\labelsep}{0em}
        \setlength{\itemindent}{-\bibhang}
        \setlength{\itemsep}{0pt}
        \setlength{\parsep}{0pt}
        \setlength{\leftmargin}{\bibhang}}
    \def\newblock{\hskip .11em plus .33em minus .07em}
    \sloppy\clubpenalty4000\widowpenalty4000
    \sfcode`\.=1000\relax}}

\newlength{\bibhang}
\let\@internalcite\cite
\def\cite{\let\@citeleft(\let\@citeright)%
    \@ifstar{\citeyear}{\citefull}}
\def\citenp{\let\@citeleft\relax\let\@citeright\relax
    \@ifstar{\citeyear}{\citefull}}
\def\citefull{\def\astroncite##1##2{##1~##2}\@internalcite}
\def\citeyear{\def\astroncite##1##2{##2}\@internalcite}

\def\@citex[#1]#2{\if@filesw\immediate\write\@auxout{\string\citation{#2}}\fi
  \def\@citea{}\@cite{\@for\@citeb:=#2\do
    {\@citea\def\@citea{; }\@ifundefined
       {b@\@citeb}{{\bf ?}\@warning
       {Citation `\@citeb' on page \thepage \space undefined}}%
{\csname b@\@citeb\endcsname}}}{#1}}

\def\@cite#1#2{\@citeleft#1\if@tempswa , #2\fi\@citeright}
\def\@biblabel#1{}

\makeatother

\def\gsim{\;\rlap{\lower 2.5pt
 \hbox{$\sim$}}\raise 1.5pt\hbox{$>$}\;}
\def\lsim{\;\rlap{\lower 2.5pt
   \hbox{$\sim$}}\raise 1.5pt\hbox{$<$}\;}

\def\angle0{\mbox{\boldmath$\theta_{\rm planet}$}}

\def\Deltabeta{\mbox{\boldmath$\Deltabeta$}}

\begin{document}

\title{How Neutral is the Intergalactic Medium at $z \sim 6$?}
\author{Adam Lidz$^{a}$, Lam Hui$^{a}$, Matias Zaldarriaga$^{b,c}$ and Roman Scoccimarro$^{b}$
\\
\vspace{0.2cm}
$^{a}$ Department of Physics, Columbia University, 538 West 120th Street,
New York, NY 10027\\   
$^{b}$ Physics Department, New York University, 4 Washington Place, New York, NY 10003\\
$^{c}$ Institute for Advanced Study, Einstein Drive, Princeton NJ 08540\\
{\tt lidz@astro.columbia.edu, lhui@astro.columbia.edu, matiasz@physics.nyu.edu, scoccima@physics.nyu.edu}
}

\begin{abstract}
Recent observations of high redshift quasar spectra reveal long gaps with
little flux.
A small or no detectable flux does not by itself
imply the intergalactic medium (IGM) is neutral. Inferring the average neutral
fraction from the observed absorption requires assumptions about
clustering of the IGM,
which the gravitational instability model supplies.  
Our most stringent constraint on the neutral fraction at $z \sim 6$ is 
derived from the mean Lyman-beta transmission measured from the $z=6.28$ 
SDSS quasar of Becker et al. -- the neutral hydrogen fraction at
mean density has to be larger than $4.7 \times 10^{-4}$.
This is substantially higher than the
neutral fraction of $\sim 3-5 \times 10^{-5}$ at $z = 4.5 - 5.7$, 
suggesting that dramatic changes take place around or just before
$z \sim 6$, even though current constraints are still consistent
with a fairly ionized IGM at $z \sim 6$. 
These constraints translate also into constraints
on the ionizing background, subject to uncertainties in the IGM temperature.
An interesting alternative method to constrain the neutral fraction is
to consider the probability of having many consecutive
pixels with little flux, which is small unless the neutral fraction is high.
It turns out that this constraint is slightly weaker than the one obtained from
the mean transmission.
We show that while the derived neutral fraction at a 
given redshift is sensitive to the power spectrum normalization, 
the size of the jump around $z \sim 6$
is not. We caution that the main systematic uncertainties include spatial 
fluctuations in the ionizing background, and the continuum placement. 
Tests are proposed. In particular, the sightline to sightline 
dispersion in mean transmission might provide a useful diagnostic. 
We express the dispersion in terms of the transmission power spectrum, 
and develop a method to calculate the dispersion for spectra that are
longer than the typical simulation box.
\end{abstract}


\section{Introduction}
\label{intro}

Recent spectroscopic observations of $z \gsim 4.5$ quasars discovered
by the Sloan Digital Sky Survey (SDSS) have opened up new windows into
the study of the high redshift intergalactic medium (IGM)
(Fan et al. 2000, Zheng et al. 2000, Schneider et al. 2001, Anderson et al. 2001
Fan et al. 2001a, Becker et al. 2001, Djorgovski et al. 2001). 
In particular, Becker et al. (2001) observed Gunn-Peterson troughs (Gunn \& Peterson 1965)
in the spectrum of a $z = 6.28$ quasar, which were interpreted as
suggesting that the universe was close to the reionization
epoch at $z \sim 6$. 

That the absorption increases quickly with redshift is not by itself surprising: 
ionization equilibrium tells us that the neutral hydrogen density is proportional
to the gas density squared, which is proportional to $(1+z)^6$ at the cosmic mean.
The evolution of the ionizing background and gas temperature will modify this
redshift dependence, but the rapid evolution of absorption remains a robust outcome.
What is interesting, as Becker et al. (2001) emphasized, is that
the observed mean transmission at redshift $z \sim 6$ is lower than what
one would expect based on an extrapolation of the column density distribution
and its redshift evolution (number density of clouds scaling as
$\sim (1+z)^{2.5}$) from lower redshifts. 
On the other hand, the popular gravitational instability theory of structure formation 
provides detailed predictions for how the IGM should be clustered, and how
this clustering evolves with redshift, which has been shown to be
quite successful when compared with observations at $z\sim 2 - 4$ (See
e.g. Cen et al. 1994, Zhang et al. 1995, Reisenegger \& Miralda-Escud\'e 1995, 
Hernquist et al. 1996, Miralda-Escud\'e 1996, Muecket et al. 1996, 
Bi \& Davidsen 1997, Bond \& Wadsley 1997, Hui et al. 1997, Croft
et al. 1998, Theuns et al. 1999, Bryan et al. 1999, McDonald et al. 2000a). 
These predictions allow us to directly infer the neutral fraction of
the IGM from the observed absorption (the relation between the two
depends on the nature of clustering of the IGM), and so 
can further inform our interpretations of the recent $z \sim 6$ results.

How neutral is the IGM at $z \sim 6$, and how different is the neutral fraction
compared to lower redshifts? These are the questions we would like to address quantitatively, 
making use of the gravitational instability model of the IGM.

The paper is organized as follows. First, we start with
a brief description of the gravitational instability model for the IGM 
and the simulation technique in \S \ref{method}. 
In \S \ref{meanTLya}, we derive the neutral hydrogen fraction $X_{\rm HI}$, 
and equivalently the level of ionizing flux $J_{\rm HI}$, at several different redshifts
leading up to $z \sim 6$ from the observed mean Lyman-alpha (Ly$\alpha$) transmission. 
This exercise using the Ly$\alpha$ spectrum
is similar to the one carried out in McDonald \& Miralda-Escud\'e
(2001), except for the addition of new high redshift data.  
\footnote{This part of the calculation involving the matching of
the mean Ly$\alpha$ transmission is also similar to a number
of earlier papers where the primary quantity of interest is the baryon density
(e.g. Rauch et al. 1997, Weinberg et al. 1997, Choudhury et al. 2000, 
Hui et al. 2001).
Here, we fix the baryon density and study the ionizing background
or the neutral fraction instead (see \S \ref{method}).}
We then examine in \S \ref{lyman beta}
the constraints on the same quantities
$X_{\rm HI}$ and $J_{\rm HI}$ from the observed mean Lyman-beta (Ly$\beta$) transmission,
Ly$\beta$ being particularly useful at high Ly$\alpha$ optical depth, because
the Ly$\beta$ absorption cross-section is a factor of $\sim 5$ smaller than the
Ly$\alpha$ cross-section. The goal here is to use
Ly$\beta$ absorption to obtain constraints on $X_{\rm HI}$ and $J_{\rm HI}$
that are as stringent as possible. 
In \S \ref{lyman beta}, we also examine the sensitivity of our conclusions
to the power spectrum normalization. 

An intriguing question is: instead of focusing on the mean transmission, 
can one make use of the fact that the observed spectrum at $z \sim 6$ contains
a continuous and long stretch ($\sim 200 - 300 \AA$) with little or no detected flux
to obtain more stringent limits on the neutral fraction or $J_{\rm HI}$? 
The idea is that since the IGM gas density naturally fluctuates spatially, 
it seems a priori unlikely to have no significant upward
fluctuation in transmission for many pixels in a row -- unless of course
the neutral fraction $X_{\rm HI}$ is indeed quite high. 
We will show in \S \ref{los constraint} this provides constraints that are
slightly weaker to those obtained using the mean transmission.

In all the simulations discussed in this paper, the ionizing background
is assumed uniform spatially, just as in the majority of high redshift IGM
simulations. A natural worry is that as the universe becomes more neutral
at higher redshifts, the ionizing background would be more non-uniform.
One way to test this is to use several lines of sight, available at
$z \sim 5.5$, and compare the observed line-of-sight scatter in
mean transmission against the predicted scatter based on simulations
with a uniform background. We discuss this in \S \ref{variance of mean t}, 
estimate the level of ionizing background fluctuations, and make predictions
for the scatter at $z \sim 6$. Here, we also introduce a
technique to handle the problem of limited box-size. 

Readers who are not interested in details can skip to \S \ref{conclusions} where
we summarize the constraints obtained. We also discuss in \S \ref{conclusions}
the issue of continuum placement, and how the associated uncertainties can
be estimated. While the work described in this paper was being carried out, several 
papers appeared which investigate related issues 
(Barkana 2001, Razoumov et al. 2001, Cen \& McDonald 2001, Gnedin 2001, Fan et al. 2001b). 
Where there is overlap, our results are in broad
agreement with these papers.  We present a comparision with other authors
at the end of \S \ref{lyman beta}.
Our approach here is most similar to that of Cen \& McDonald (2001). 
In addition to obtaining constraints on the ionizing background 
from the Ly$\alpha$ and Ly$\beta$ transmission as was considered by
Cen \& McDonald, 
we consider the possible constraint from the Gunn-Peterson trough itself, 
examine the dependence on power spectrum normalization, and
develop a method to predict the scatter in mean transmission
by relating it to the power spectrum, which might be of wider interest. 
We also place a slightly stronger emphasis on the neutral fraction, which is more
robustly determined compared to the ionizing background or photoionization
rate. 

\section{The Gravitational Instability Model for the IGM}
\label{method}

The Ly$\alpha$ optical depth is related to the IGM density, assuming ionization equilibrium
, via
\begin{equation}
\tau_\alpha = A_\alpha (1+\delta)^{2 - 0.7 (\gamma - 1)}
\label{taualpha}
\end{equation}
where $\delta$ is the gas overdensity ($\delta = (\rho - \bar\rho)/\bar\rho$, where
$\rho$ is the gas density and $\bar \rho$ its mean), $\gamma$ is the equation
of state index for the IGM\footnote{The photoionized IGM at overdensity of a few or less
is expected to follow a tight temperature-density relation of the form
$T = T_0 (1+\delta)^{\gamma-1}$, where $T$ is the gas temperature and $T_0$ is
its value at the cosmic mean density (see Hui \& Gnedin 1997). We caution that close to reionization, these quantities may not be a function of $\delta$
alone.  The IGM may be heated inhomogenously, causing spatial fluctuations
in $T_0$ and $\gamma$.}, and $A_\alpha$ is
given by (see e.g. Hui et al. 2001 and references therein):

\begin{eqnarray}
\label{Aalphanew}
A_\alpha = 51 \left[X_{\rm HI} \over 1.6 \times 10^{-4}\right] 
\left[\Omega_b h^2 \over 0.02\right] \left[{0.65 \over h}\right]
\left[{1+z \over 7}\right]^3 \left[{11.7 \over H(z)/H_{0}}\right]
\end{eqnarray}
where $X_{\rm HI} \equiv n_{\rm HI} / n_{\rm H}$ 
($n_{\rm H}$ is the total density of neutral and ionized hydrogen, and
$n_{\rm HI}$ is the neutral hydrogen density) is the neutral hydrogen fraction at mean density 
($\delta = 0$).
\footnote{The neutral fraction at arbitrary $\delta$ is given by $X_{\rm HI}$ times
$(1+\delta)^{1 - 0.7(\gamma-1)}$. Throughout this paper, 
whenever we quote values for $X_{\rm HI}$,
we refer to the neutral hydrogen fraction at the cosmic mean density
$\delta = 0$. }
Here $H(z)$ is the Hubble parameter at redshift $z$, $H_{0}$ is the Hubble
parameter today, $H_{0} = 100 h$ km/s/Mpc, $\Omega_{b}$ is the
baryon density in units of the critical density. 
The value of $11.7$ for $H(z)/H_{0}$ above corresponds
to that appropriate for
a cosmology with $\Omega_{m}=0.4$ and $\Omega_{\Lambda}=0.6$ at $z=6$, where
$\Omega_{m}$ and $\Omega_{\Lambda}$ are the matter and vacuum densities
in units of the critical density today. 

The neutral fraction $X_{\rm HI}$ is related to the ionizing background by
\footnote{This
equation assumes that hydrogen is highly ionized and that helium is largely doubly 
ionized.  If helium is only singly ionized, the relation between $J_{\rm HI}$
and $X_{\rm HI}$ will be changed slightly: the right hand side 
of eq (\ref{XHI}) will be multiplied by $0.93$. 
}
\begin{equation}
\label{XHI}
X_{\rm HI} = 1.6 \times 10^{-4} 
\left[{\Omega_b h^2 \over 0.02}\right]
\left[2.55 \times 10^{-2} \over J_{\rm HI}\right]
\left[{T_0 \over 2 \times 10^4 {\, \rm K}}\right]^{-0.7}
\left[{1+z \over 7}\right]^3 
\end{equation}
where the dimensionless quantity 
$J_{\rm HI}$ is related to the photoionization rate $\Gamma_{\rm HI}$ by
\begin{equation}
\Gamma_{\rm HI} = 4.3 \times 10^{-12} J_{\rm HI} {\, \rm s^{-1}}
\label{Gamma} 
\end{equation}
The quantity $J_{\rm HI}$ provides a convenient way of describing the normalization of
the ionizing background, without specifying the exact spectrum, in a
way that is directly related to the physically relevant $\Gamma_{\rm HI}$
(e.g. Miralda-Escud\'e et al. 1996). 
It is related to the specific intensity at $912 \AA$ $j_{\nu_{\rm HI}}$ by 
$J_{\rm HI} = j_{\nu_{\rm HI}} \times [3 / (\beta + 3)]$, where
$\beta$ is the slope of the specific intensity just blueward
of $912 \AA$ ($j_{\nu} 
\propto \nu^{-\beta}$ where $\nu$ is frequency), and $j_{\nu_{\rm HI}}$ is 
measured in the customary units of
$10^{-21} {\,\rm erg \, s^{-1} \, cm^{-2} \, Hz^{-1} \,
ster^{-1}}$ (for non-power law $j_{\nu}$, eq. [\ref{Gamma}] provides
the exact definition for $J_{\rm HI}$; see e.g. 
Hui et al. 2001).

Two more ingredients should be mentioned to complete the specification of our model
for the Ly$\alpha$ absorption (see e.g. Hui et al. 1997 for details). First, the optical depth as a function of
velocity is computed by taking the right hand side of eq. (\ref{taualpha})
in velocity space (i.e. taking into account peculiar velocities) and smoothing
it with a thermal broadening window. Second, the gas density and velocity
fields are predicted by some Cold Dark Matter (CDM) cosmological model using
numerical simulations. 

There are obviously a number of free parameters in our model. Let us
discuss each of them in turn. 

Throughout this paper, we assume $\Omega_b h^2 = 0.02$, as supported
by recent cosmic microwave background measurements 
(Netterfield et al. 2001, Pryke et al. 2001) 
and the nucleosynthesis constraint from primordial 
deuterium abundance (Burles et al. 2001). 
We also assume throughout $h = 0.65$, $\Omega_m = 0.4$, and $\Omega_\Lambda = 0.6$. 
Variations of these parameters within the current bounds do not 
contribute significantly to the uncertainties of the constraints obtained in
this paper (see Hui et al. 2001). 

The temperature $T_0$ and equation of state index $\gamma$ at the
redshifts of interest in this paper are somewhat uncertain.
There are no direct measurements of the thermal state of the IGM
at our redshifts of interest, $z \gsim 4$. 
Measurements at $z \lsim 4$ yield values consistent with 
$T_0 = 2 \times 10^4$ K and $\gamma = 1$ (McDonald et al. 2000b,
Ricotti et al. 2000, Zaldarriaga et al. 2001.  
Schaye et al. 2000, however, measure a slightly
lower temperature).
Given that the temperature right after reionization is expected
to be about $25000$ K with $\gamma = 1$ (with some dependence on the 
hardness of the ionizing spectrum; see e.g. Hui \& Gnedin 1997), 
which is not too different from the measurements at $z \lsim 4$, 
we will assume throughout this paper, when making use of eq.
(\ref{XHI}) to infer $J_{\rm HI}$, that $T_0 = 2 \times 10^4$ K and $\gamma = 1$. 
Note that while the theoretically allowed range for $\gamma$ is from
$1$ to $1.6$ (Hui \& Gnedin 1997), what matters for our purpose
is $2 - 0.7 (\gamma - 1)$ (eq. [\ref{taualpha}]), which only ranges
from $1.58$ to $2$, and does not significantly affect our results.
It is also important to emphasize that {\it the inference of
$X_{\rm HI}$ from observations, unlike the case for $J_{\rm HI}$, is
not subject directly to uncertainties in the temperature $T_0$}. This is because observations
constrain $A_\alpha$ from which we can obtain $X_{\rm HI}$ without
knowing $T_0$ (see eq. [\ref{Aalphanew}]).
\footnote{
The above statement is subject to two small caveats. 
First, the optical depth given in eq. (\ref{taualpha}) has to be smoothed
with a thermal broadening window whose width depends on $T_0$. We
find that in practice, the exact width of the thermal broadening
kernel does not affect very much quantities such as the mean transmission,
which is what we will be interested in. 
Second, $T_0$ also affects the gas dynamics via the pressure
term in the equation of motion. As we will discuss below,
the effect of varying $T_0$ also appears to be small in this regard.}

To generate realizations of the density and velocity fields for a given cosmology, we 
run Hydro-Particle-Mesh (HPM) simulations (Gnedin \& Hui 1998).
The HPM algorithm is essentially a Particle-Mesh code, modified to incorporate
a force term due to gas pressure in the equation of motion.
\footnote{The temperature-density relation has to be specified as a function
of redshift in the HPM code to compute the pressure term. 
We follow McDonald $\&$ Miralda-Escud\'e (2001)
and linearly interpolate between $T_0 = 19000$ K and $\gamma = 1.2$ at $z=3.9$
and $T_0 = 25000$ K and $\gamma = 1$ at the redshift of reionization $z_{\rm reion}$. 
We found that assuming $z_{\rm reion} = 7$ versus $z_{\rm reion} = 10$ 
results in negligible difference in our results, in particular concerning 
the mean decrement and the probability distribution of transmission. 
All results in this paper are quoted from the $z_{\rm reion} = 7$ HPM simulations.
Note that in inferring $X_{\rm HI}$ and $J_{\rm HI}$ from 
eq. (\ref{Aalphanew}) and (\ref{XHI}), we always use $T_0 = 20000$ K and
$\gamma = 1$ for simplicity, as mentioned before.}
For the initial power spectrum, we use a Cold Dark Matter (CDM) type
transfer function, as parameterized by Ma (1996), which is very similar
to the commonly used Bardeen et al. (1986) transfer function. 
For the primordial spectral slope, we adopt 
$n = 0.93$ (Croft et al. 2000, McDonald et al. 2000a). 
For the linear power spectrum normalization, we employ the range
suggested by measurements from the Ly$\alpha$ forest of Croft et al. (2000):
$\Delta^2 (k) \equiv 4 \pi k^3 P(k) / (2\pi)^3 
= 0.74^{+0.20}_{-0.16}$ at $z = 2.72$ at a velocity scale
of $k = 0.03 ({\,\rm km/s})^{-1}$.\footnote{This normalization corrects
an error in an earlier draft of Croft et al. (2000). (R. Croft, private
communication.)} 
We, however, caution that the error-bar given is somewhat dependent
on the assumed error of the mean transmission measurements, which
is sensitive to the accuracy of the continuum-fitting procedure
(see e.g. Zaldarriaga et al. 2001  for a slightly different assessment of the error-bar).  
The power spectrum in this model has a similar shape to that of
favored cosmological models, but slightly lower amplitude (Croft et al. 2000).
In \S \ref{lyman beta} we demonstrate that our main conclusion, that the 
neutral fraction increases dramatically near $z \sim 6$, is insensitive to
our assumptions about the amplitude of the power spectrum.
In practice, we examine models with 
different normalizations by running a simulation with outputs at
several different redshifts: each redshift then corresponds to a different
power spectrum normalization, and linear interpolation is performed to
reach any desired normalization. 
\footnote{We do not vary the primordial spectral index $n$ here. Quantities such as the
mean transmission which we are interested in here are generally sensitive to power on only 
a small range of scales. 
Varying $n$ is therefore largely degenerate with varying $\Delta^2$.}

Our simulations have a box size of $8.9$ Mpc/h, with a $256^3$ grid.
McDonald \& Miralda-Escude (2001) found this box size and resolution to
be adequate for IGM studies up to $z \sim 5$. We have verified that
the same is true up to $z = 6$, in the sense that the transmission probability
distribution has converged for our choice of simulation size and resolution.

Finally, we should say a few words about simulations of the Ly$\beta$ region.
In regions of the quasar spectrum that are between $973 \AA (1 + z_{\rm em.})$
 and $1026 \AA (1 + z_{\rm em.})$, where
$z_{\rm em.}$ is the redshift of the quasar, two kinds of absorption
can exist: one is Ly$\beta$ due to material at redshift 
$ 0.948 (1+ z_{\rm em.})  < 1+z < 1+z_{\rm em.}$, the other is
Ly$\alpha$ due to material at redshift 
$ 0.800 (1 + z_{\rm em.}) <  1 + z < 0.844 (1+ z_{\rm em.})$. 
In other words, in such a region, the observed optical depth
would be given by $\tau = \tau_\beta + \tau_\alpha$
where $\tau_\beta$ and $\tau_\alpha$ arises at different redshifts.
The Ly$\alpha$ optical depth can be computed as before.
The Ly$\beta$ optical depth $\tau_\beta$ can be computed using 
eq. (\ref{taualpha}), except that $A_\alpha$ is replaced by $A_\beta$:
\begin{equation}
A_\beta = {1 \over 5.27} A_{\alpha}
\label{Abeta}
\end{equation}
The factor of $5.27$ reflects the fact that the Ly$\beta$ transition
has a cross-section that is $5.27$ times smaller than 
Ly$\alpha$.

\section{Constraints on the Neutral Hydrogen Fraction
and the Ionizing Background}
\label{Ionizing Background}

\subsection{Constraints from the Ly$\alpha$ Mean Transmission}
\label{meanTLya}

Using eq. (\ref{taualpha}) and (\ref{Aalphanew}), we compute the $X_{\rm HI}$, which
also fixes $J_{\rm HI}$ (eq. [\ref{XHI}]), 
necessary to match the observed Ly$\alpha$ mean transmission
$\langle e^{-\tau_\alpha} \rangle$
at $z = 4.5 - 6$ (see Table 1 for a summary of the measurements). 
The results of our calculation are presented in Fig. \ref{meanTalpha}. 
This plot also contains a point at $z=6.05$ which is the result of matching
the mean transmission in the Ly$\beta$ forest, as we describe in \S 
\ref{lyman beta}.

Also shown in the figure is a dotted line which shows
$X_{\rm HI} \propto (1+z)^{3}$, which appears to be a good fit to
the data from $z = 4.5$ to $z=5.7$. From eq. (\ref{XHI}), one can see that
such a trend for the neutral fraction 
is equivalent to assuming constant $J_{\rm HI}$ (or more
accurately, constant $J_{\rm HI} T_0^{0.7}$; see eq. [\ref{XHI}]). 

As one can see, ignoring for now the Ly$\beta$ point, 
the neutral fraction does appear to have a modest jump 
around $z \sim 6$: it increases by a factor of $\sim 4.0$ from $z=5.7$ to 
$z=6.05$, while it changes at most by $\sim 1.9$ from $z=4.5$ to $z=5.7$. 
A similar trend (but opposite in sign) 
can be seen in the ionizing flux $J_{\rm HI}$. 
The 1 $\sigma$ error-bar here takes into account the measurement error in
mean transmission, and the range of power spectrum normalization
stated in \S \ref{method}. As we have explained in \S \ref{method},
while $X_{\rm HI}$ is not sensitive to the assumed temperature 
of the IGM ($T_0$), our constraints on $J_{\rm HI}$ are directly influenced
by it. As emphasized before, we assume $T_0 = 2 \times 10^4$ K. 
In other words, our constraints on $J_{\rm HI}$ are really constraints
on the quantity $J_{\rm HI} (T_0 / 2 \times 10^4 {\,\rm K})^{0.7}$
(see eq. [\ref{XHI}]). It is therefore straightforward
to rescale our constraints on $J_{\rm HI}$ if the temperature were a little
bit different. \footnote{The temperature also affects the thermal broadening
window, but we find that in practice its effect on our constraints on
$A_\alpha$ (eq. [\ref{Aalphanew}]) is small; see discussion in \S \ref{method}.}
It is an interesting question to ask whether the apparent jump in the ionizing
flux can instead be attributed to a jump in the temperature. 
In general, the temperature is expected to evolve slowly with redshift
after reionization (Hui \& Gnedin 1997). 

Regarding the measurement error, we should also emphasize that
the Becker et al.'s 2 $\sigma$ error actually includes the possibility
of having zero transmission at $z = 6.05$. This means that at 2 $\sigma$,
we would only have a lower limit on $X_{\rm HI}$, or 
an upper limit on $J_{\rm HI}$, for the highest redshift point in Fig. \ref{meanTalpha},
allowing the possibility that the IGM is neutral at $z \sim 6$, $X_{\rm HI} = 1$.

\subsection{Constraints from the Ly$\beta$ Mean Transmission}
\label{lyman beta}

In this section we consider the constraints placed by Becker et al.'s
measurement of the mean transmission in the Ly$\beta$ region.
Absorption in the Ly$\beta$ region has two components: $\tau = \tau_\alpha + \tau_\beta$,
where the Ly$\alpha$ optical $\tau_\alpha$ and 
the Ly$\beta$ optical depth $\tau_\beta$ originate at different redshifts. 
Ly$\beta$ absorption due to material 
at $z = 6$ coincides in wavelength with Ly$\alpha$ absorption due to 
material at $z = (1+6) \times 1026/1216 - 1 = 4.9$. Because the points
of origin are so widely separated, they can be effectively treated as 
statistically independent i.e. $\langle e^{-\tau} \rangle = 
\langle e^{-\tau_\alpha} \rangle \langle e^{-\tau_\beta} \rangle$. 
Becker et al. measured $\langle e^{-\tau_{\rm \beta}} \rangle$ at 
$z \sim 6$ by dividing the net mean transmission $\langle e^{-\tau} \rangle$
in the Ly$\beta$ region by the mean transmission in Ly$\alpha$ at $z \sim 5$. They
obtained $\langle e^{-\tau_{\beta}} \rangle = -0.002 \pm 0.020$.
Clearly, this measurement is consistent with a completely neutral IGM.
However, the interesting question is: what kind of lower limit does it
set on the neutral fraction, and does it improve upon the lower limit 
from the mean Ly$\alpha$ absorption ?

We carry out a calculation that is analogous to what is
described in \S \ref{meanTLya}, except for the key difference
that in computing $\tau_\beta$ from the simulated density and velocity fields,
we employ $A_\beta$ which is a factor of $5.27$ smaller than $A_\alpha$
(see eq. [\ref{taualpha}] \& [\ref{Abeta}]). 
The results of our calculation are shown in Fig. \ref{meanTalpha}, as the
highest redshift points in the plot, which have error bar arrows 
pointing towards
a completely neutral IGM and a vanishing ionizing background. It can be seen 
that the ($1-\sigma$) lower limit
on $X_{\rm HI}$ is $X_{\rm HI} > 4.7 \times 10^{-4}$.  This is a 
factor of $\sim 3$ larger than the neutral fraction required to match
the upper limit on the mean transmission in Ly$\alpha$ for our 
fiducial normalization, and a slightly stronger constraint 
than that obtained in section 
\ref{meanTLya}, including the uncertainty in power spectrum normalization. 
Similarly, the upper limit on $J_{\rm HI}$ is 
$J_{\rm HI} < 9.0 \times 10^{-3}$.
The moral here is that because the Ly$\beta$ absorption cross-section is 
a factor of $5.27$ smaller than the Ly$\alpha$ cross-section, Ly$\beta$
offers a more sensitive probe of the neutral fraction, especially when
the Ly$\alpha$ optical depth is high. 

The neutral fraction at $z \sim 6$ is thus a factor of $\sim 10$ higher than
that at redshift $z \sim 5.7$, where it is $X_{\rm HI} = 4.9 \times 10^{-5}$.
This dramatic change in the neutral fraction is suggestive, probably indicating
that the reionization epoch is nearby.

Furthermore, this conclusion is not sensitive to our assumptions about
the amplitude of the power spectrum. 
Although the neutral fraction at redshift $z=6.05$ is itself sensitive to the
amplitude of the power spectrum, we find that the factor by which the 
neutral fraction increases from $z=5.7$ to $z=6.05$ depends only weakly
on the amplitude.  In Fig. \ref{xbreak} we plot both the neutral
fraction at $z=6.05$ and the jump in the
neutral fraction for a range of different power spectrum normalizations.
The jump is defined as the ratio $X_{\rm HI} (z=6.05) / X_{\rm HI} (z=5.7)$.
Here $X_{\rm HI} (z=6.05)$ is the lower limit resulting from the 1 $\sigma$
error in the mean transmission in Ly$\beta$ at $z=6.05$ and the 
error bars in the jump arise from the 1 $\sigma$ error in the mean 
transmission in Ly$\alpha$ at $z=5.7$.  As one can see in the plot,
the lower limit on the neutral fraction at $z=6.05$ varies from 
$X_{\rm HI} > 3 \times 10^{-4}$ to $X_{\rm HI} > 9 \times 10^{-4}$ as
$\Delta^{2}(k=0.03 ({\,\rm km/s})^{-1}, z=2.72)$ varies from $0.5$ to $1.3$.
The neutral fraction itself varies significantly with 
power spectrum normalization,
scaling approximately as $X_{\rm HI} \propto 
[\Delta^{2}(k=0.03 {\,\rm s/km}, z=2.72)]^{1.1}$, for this range of
normalizations.
The jump, however, changes only slightly over a large range of normalizations.
As $\Delta^{2}(k=0.03 ({\,\rm km/s})^{-1}, z=2.72)$ varies from $0.5$ to $1.3$,
the jump changes only from $\sim 9.7$ to $\sim 11.1$.
Our conclusion that the neutral
fraction of the IGM increases dramatically near $z \sim 6$ seems robust.

One can also consider the absorption in the Ly$\gamma$ region, or
even the higher Lyman series. In practice, the accumulated amount
of absorption from Ly$\alpha$ as well as Ly$\beta$ at different redshifts
makes it harder to measure the Ly$\gamma$ transmission itself with good
accuracy. 

Our constraints on the neutral fraction and the intensity of the
ionizing background are
consistent with those found by other authors, given our different choices of
power spectrum normalization.  Fan et al. (2001b) found, from the mean
Ly$\beta$ transmission, that
$\Gamma_{\rm -12} < 0.025$, where $\Gamma_{\rm -12}$ is the photoionization
rate of eq (\ref{Gamma}) in units of $10^{-12} {\, \rm s^{-1}}$.  Although
this constraint is somewhat stronger than the constraint implied by our 
fiducial model,
$\Gamma_{\rm -12} < 0.039$, we expect the difference due to our different 
power spectrum normalizations.
Fan et al.'s (2001b) constraint comes from semi-analytic arguments, 
shown consistent with an LCDM simulation with $\Omega_{m}=0.3$, 
$\Omega_{\Lambda}=0.7$,$h=0.65$,
$\Omega_b h^2 = 0.02$, and $\sigma_{8}=0.9$.  This model has a substantially 
larger normalization, $\Delta^{2}(k=0.03 ({\,\rm km/s})^{-1}, z=2.72)=1.25$,
than our fiducial model of $\Delta^{2}(k=0.03 ({\,\rm km/s})^{-1}, z=2.72)=
0.74$.  The difference in normalization reflects some tension between the 
normalization derived from the observed cluster abundance, which Fan uses, 
and that from the Lyman-$\alpha$ forest which our model is based on 
(Croft et al. 2000).  
Fan et al. (2001b) assume $T_0 = 2.0 \times 10^4$ K in
placing their constraint. Their limit, $\Gamma_{\rm -12} < 0.025$, 
includes only uncertainties in the
mean transmission and not additional uncertainties from the power spectrum
normalization.   From figure (\ref{xbreak}), we infer that 
Fan et al.'s (2001b) normalization 
implies $X_{\rm HI} > 8.8 \times 10^{-4}$ in our cosmology.  
Rescaling this result from our 
assumed $\Omega_{m}=0.4$ to an $\Omega_{m}=0.3$ cosmology, and using eqs.
(\ref{XHI}) and (\ref{Gamma}) we predict $\Gamma_{\rm -12} < 0.024$, or
$X_{\rm HI} > 7.6 \times 10^{-4}$, for
Fan et al.'s (2001b) model.  The constraint of Fan et al. (2001b) 
is thus consistent with our constraint given our 
different choice of normalization.  Cen \&
McDonald (2001), using a model similar to that of Fan et al. (2001b), 
obtained the constraint $\Gamma_{\rm -12} < 0.032$, using the 
Ly$\beta$ mean transmission.  The constraint is slightly
weaker than that of Fan et al. (2001b) and our extrapolation to their 
normalization, because Cen \& McDonald (2001) consider a larger upper limit
to the observed mean transmission, including an estimate of
the uncertainty due to sky subtraction.  At
slightly lower redshifts, we can also compare with the results of
McDonald \& Miralda-Escud\'e (2001) derived from matching the mean Ly$\alpha$
transmission.  For example, at $z=5.2$, these
authors found $\Gamma_{\rm -12}=0.16$ to match the observed mean transmission
of $\langle e^{-\tau_{\beta}} \rangle = 0.09$.  McDonald \& Miralda-Escud\'e
(2001) consider a model whose normalization we infer to be  
$\Delta^{2}(k=0.03 ({\,\rm km/s})^{-1}, z=2.72)=0.98$.  To match the same
mean transmission with this normalization we infer a somewhat 
higher photoionization rate, $\Gamma_{-12}=0.19$.  Part of the difference
may be that the $\Gamma_{\rm -12}$ necessary to match a given mean 
transmission 
varies by $\sim 5\%$ between two different realizations of the 
density field.  The remaining difference may come from the procedure of 
linearly interpolating
between outputs or from some modeling difference.  At any rate, our
results are roughly consistent with those of other authors given our different power spectrum normalizations.

\subsection{Constraints from the Gunn-Peterson Trough Itself -- the Fluctuation Method}
\label{los constraint}

The fact that Becker et al. (2001) observed a Gunn-Peterson trough, where
a long stretch of the spectrum contains little or no flux, can conceivably
be used to further tighten the constraints obtained from the previous
sections. Since the IGM is expected to have spatial fluctuations, the
probability of having many pixels in a row turning up a very small transmission
must be low, unless the neutral fraction is intrinsically quite high. 
The same reasoning can be applied to either the Ly$\alpha$ or Ly$\beta$
absorption. We will discuss our method for Ly$\alpha$ in detail. The method
for Ly$\beta$ is a straightforward extension. For simplicity,
we will call this method, the fluctuation method.

Becker et al. (2001) finds from the spectrum of SDSS 1030+0524, 
the $z = 6.28$ quasar, the Ly$\alpha$ transmission is consistently below
about 0.06 for a region that spans $260 \AA$, between $8450 \AA$ to $8710 \AA$. 
The noise level per $4 \AA$ pixel is $\sqrt{\langle n^2 \rangle} \sim 0.02$,
where $n$ represents the photon noise fluctuation.
\footnote{We estimate the noise per pixel from Becker et al.'s 
error-bar in the mean transmission, which is $\sim 0.003$.
This is estimated from a chunk of the spectrum which is $260 \AA$ long,
and so the dispersion per $4 \AA$ pixel should be approximately
$\sqrt{\langle n^2 \rangle} \sim 0.003 \times \sqrt{65} \sim 0.02$. 
Note that the actual dispersion varies across the spectrum, but this
should suffice as a rough estimate. This estimate also agrees with
an estimate of the error by comparing Fig. 1 and Fig. 3 of Becker et al.}
The observed transmission $F$ at a given pixel is
$F = e^{-\tau} + n$, where $e^{-\tau}$ is the true transmission. 
The noise here should be dominated by Poisson fluctuations of
the subtracted sky background (as well as perhaps read-out error). 
Let $P(F_1 , F_2, ... F_N) dF_1 ... dF_N$ be the probability that
$N$ consecutive pixels have observed transmission fall into the range
$F_1 \pm dF_1/2$ ... $F_N \pm dF_N/2$. In our case, $N = 65$ for
the pixel size of $4 \AA$. The problem is then to find
the probability $\int_{< 0.06} ... \int_{< 0.06} P(F_1 ... F_N) dF_1 ... dF_N$
as a function of $J_{\rm HI}$, and ask what maximal $J_{\rm HI}$
(or equivalently, minimal $X_{\rm HI}$) would give an acceptable
probability. By choosing the ``acceptable probability'' to be
within $68\%$ of the maximum likelihood 
(maximum likelihood is achieved when the neutral fraction is unity), we obtain 
the 1 $\sigma$ upper limit on 
$J_{\rm HI}$ or 1 $\sigma$ lower limit on $X_{\rm HI}$.

Our simulation has a comoving box size of $8.9$ Mpc/h, corresponding 
to $42 \AA$ for Ly$\alpha$ at $z \sim 6$, which falls short of the wavelength
range we need for this problem, which is $260 \AA$. 
In other words, the probability $P(F_1 , F_2, ... F_N) dF_1 ... dF_N$ can be
estimated directly from our simulation only for $N \le 10$. 
However, the mass correlation length scale at this redshift
($\lsim 1$ Mpc/h) is actually a fraction of the box size, which
means one can treat fluctuations on scales beyond the box size as
roughly uncorrelated. Assuming so, we estimate 
$\int_{< 0.06} ... \int_{< 0.06} P(F_1 , F_2, ... F_{10}) dF_1 ... dF_{10}$
from the simulation, and then take its $6$-th power, which would give
us the probability that $60$ consecutive pixels have transmission below
$0.06$. This is slightly smaller than the number $65$ that we need, but
at least will provide us conservative constraints on $X_{\rm HI}$ and $J_{\rm HI}$. 
We have also tested our approach by using fractions of the box-size as
a unit, and find that our results do not change significantly (less than $10 \%$). 

Fig. \ref{problong}a (dotted curve) shows our estimate of the probability 
$\int_{< 0.06} ... \int_{< 0.06} P(F_1 , F_2, ... F_{N}) dF_1 ... dF_{N}$
for $N = 60$ and pixel size $4 \AA$, as a function of $J_{\rm HI}$. 
Our simulated spectra have been convolved with the observation resolution
(full-width-at-half-maximum of $1.8 \AA$), rebinned into pixels of $4 \AA$ each
and added Gaussian noise with a dispersion of $0.02$. 
From the dotted curve in Fig. \ref{problong}a, applying a likelihood analysis, we obtain a
1 $\sigma$ upper limit on $J_{\rm HI}$ of $J_{\rm HI} < 0.014$, and
a corresponding lower limit on $X_{\rm HI}$ of $X_{\rm HI} > 2.95 \times 10^{-4}$. 
This is for a model with a power spectrum normalization of
$\Delta^2 (k = 0.03 {\,\rm s/km}, z = 2.72) = 0.74$ (see \S \ref{method}). 
The mean Ly$\alpha$ transmission constraints for the same model are
$J_{\rm HI} < 0.028$ and $X_{\rm HI} > 1.5 \times 10^{-4}$. 
\footnote{Do not confuse these constraints, which are for the
particular power spectrum normalization mentioned above, to
the constraints discussed in earlier sections, which include the
uncertainty in the power spectrum normalization. We focus
on a single model in this section for simplicity.}
{\it This means that considering Ly$\alpha$ alone, the fluctuation method
yields somewhat stronger constraints
compared to using simply the mean transmission. }

Fig. \ref{problong}b (dotted curve) shows the same methodology applied
to the Ly$\beta$ Gunn-Peterson trough. 
A new ingredient here is that one needs an additional simulation
of the same model at redshift $z = 4.9$ to produce the Ly$\alpha$
absorption that can be overlaid on top of the Ly$\beta$ absorption
from $z = 6.05$. This additional simulation should have different initial phases
to mimic the fact that fluctuations at $z = 4.9$ and those at 
$z = 6.05$ should be uncorrelated. 
We obtain 1 $\sigma$ limits
of $J_{\rm HI} < 0.012$ and $X_{\rm HI} > 3.4 \times 10^{-4}$. This
is about $40 \%$ weaker than the constraints we obtain from the
Ly$\beta$ mean transmission. {\it In other words, from Ly$\beta$ absorption,
the fluctuation method yields slightly weaker constraints
compared to using the mean transmission.}  It is also only slightly stronger
than the constraint obtained from the fluctuation method applied to 
Ly$\alpha$.

It is an interesting question to ask how many sightlines one would
need to improve the constraints by, say a factor of $2$. Our
approach can be easily extended to multiple (uncorrelated) sightlines,
and we find that about $5$ sightlines (each containing
a Gunn-Peterson trough of the same length and
same signal to noise) are necessary for such an improvement.

Part of the difficulty with obtaining stronger constraints, in addition
to the small number of sightlines, is the dominance of noise. 
The lower panel of Fig. \ref{pdf} shows the one-pixel ($4 \AA$) 
probability distribution function (PDF) of the true 
transmission $e^{-\tau}$ (i.e. no noise added), for three different values
of $J_{\rm HI}$ (the power spectrum normalization is
the same as that in Fig. \ref{problong}). 
The upper panel shows the corresponding PDF's of the observed transmission
$F$ (i.e. after convolving $P(e^{-\tau})$ with a Gaussian of dispersion
$0.02$). As expected, noisy data make the PDF's more similar. 
Nonetheless, as we pointed out above, with sufficient
number of sightlines, there might be
a non-negligible chance of seeing pixels with high transmission that
take place at the tail of the PDF's, hence allowing us to distinguish between
the different levels of the ionizing background.
Alternatively, one can try improving the signal-to-noise per pixel.
In Fig. \ref{problong}b, we show with a dashed curve the 
corresponding probability if the noise per pixel is lowered by a factor of $4$. 
The constraints improve by a little more than a factor of $2$. 
We should emphasize, however, systematic errors are likely important here
-- we will discuss them in the next two sections.

\section{The Variance of the Mean Transmission}
\label{variance of mean t}

If, as is suggested by our discussion in
\S \ref{lyman beta} (see Fig. \ref{meanTalpha}), the IGM is close to the epoch of reionization 
at $z \sim 6$, one might expect large fluctuations in the ionizing background
near that time.
For instance, one line of sight might probe a region of the IGM where the
ionized bubbles around galaxies or quasars have percolated, while another
might probe the pre-percolation IGM. 
As mentioned before, the simulations we employ do not take into account
fluctuations in the ionizing background.  (For simulations incorporating 
radiative transfer see e.g., Gnedin \& Abel 2001, Razoumov 2001). One useful check would then
be to predict the sightline to sightline scatter in mean transmission 
from our simulations, and compare that against the observed scatter.
At $z \sim 5.5$, 4 lines of sight are available for a measurement of the scatter.
We will examine this, as well as make predictions for the scatter at $z \sim 6$,
which more high redshift quasars in the future will allow us to measure.

Our estimate relies on simulation measurements of the transmission
power spectrum. This is in contrast to an estimate of the same 
quantity made by Zuo 
(Zuo 1993) who makes a prediction based on extrapolations of the
column density distribution and of the number of absorbing clouds
per. unit redshift (Zuo \& Phinney 1993).  Zuo also assumes that the clouds
are Poisson distributed, while our measurement incorporates the clustering
in the IGM via our numerical simulation.

An immediate problem presents itself: sightlines from
which the mean transmission is measured are typically longer than the
usual simulation box. We tackle this problem by expressing the
variance of mean transmission in terms of the transmission power spectrum,
and making use of a reasonable assumption about the behavior of the
power spectrum on large scales.

The mean transmission from one sightline is estimated using
\begin{equation}
{\bar F} = {1 \over N} \sum_{i=1}^N F_i
\end{equation}
where $N$ is the number of pixels, $F_i$ is the
observed transmission at pixel $i$, $F_i = f_i + n_i$ where
$f = e^{-\tau}$ is the true transmission and $n$ is the noise fluctuation.
We use the symbol ${\bar F}$ to represent the estimator,
and $\bar f$ to denote the true mean transmission. 
The variance of the estimated mean transmission is then
\begin{eqnarray}
\label{sigmaTeq}
&& \sigma_T^2 \equiv \langle {\bar F}^2 \rangle - \langle \bar F \rangle^2 
= {1 \over N^2} \sum_{i,j} [\langle F_i F_j \rangle - \langle F_i \rangle
\langle F_j \rangle] \\ \nonumber &&
= {1\over N^2} \sum_{i,j} \xi_{ij} + {1\over N} \sigma_n^2
= 2 \int_0^\infty {dk \over 2 \pi} \left[{\,\rm sin} (k L /2) \over
kL/2 \right]^2 P_f (k) + {\sigma_n^2 \over N}
\end{eqnarray}
where $\sigma_n^2 \equiv \langle n^2 \rangle$ is assumed roughly independent
of position, and $\xi_{ij}$ is the un-normalized two-point correlation
of the transmission i.e. $\xi_{ij} \equiv \langle f_i f_j \rangle 
- {\bar f}^2$, and $P_f (k)$ is its one-dimensional Fourier
transform. The symbol $L$ denotes the comoving length of the spectrum from
which the mean transmission is measured, and $k$ is the comoving wavenumber.

To evaluate $\sigma_T$, we need to know the transmission power spectrum on scales generally larger
than the size of the typical simulation box. 
It is expected that the transmission power spectrum takes the shape (not
the normalization) of the linear 
mass power spectrum on large scales (i.e. essentially linear biasing; see 
Scherrer \& Weinberg 1998, Croft et al. 1997, Hui 1999). 
We therefore use this to extrapolate the simulation $P_f (k)$ to large scales (small $k$'s).  We find that $P_f (k)$ is well approximated by
$P_f (k) = B {\,\rm exp} (-ak^2) \int_k^\infty  (dk /2\pi) k P_{\rm mass} (k)$
where $P_{\rm mass}$ is the three-dimensional linear mass power spectrum.

Becker et al. gave an estimate of $\sigma_T \sim 0.03 \pm 0.01$,
$\bar f = 0.1$, at $z = 5.5$ using
4 different sightlines, each spanning $\Delta z = 0.2$, which
corresponds to $L \sim 57$ Mpc/h. \footnote{The error on $\sigma_T$ is estimated assuming Gaussian statistics and that the four lines of sight are 
independent.
Then $var(\sigma_T) = \sigma_T^{2} / 2 n$ (see e.g. Kendall \& Stuart 1958).}
An estimate of the noise term is provided by the error in the mean
transmission, $(\sigma_n^2 / N)^{0.5} \sim 0.003$.
For the $\Delta^2 (k = 0.03 {\,\rm s/km}, z=2.72) = 0.74$ case, the
fitting parameters are $B = 0.033$ and $a = 0.013 {\,\rm Mpc^2 / h^2}$.
Using eq. (\ref{sigmaT}), we then find $\sigma_T = 0.030$ for $\Delta^2 (k = 0.03 {\,\rm s/km}, z=2.72)=0.74$, $\sigma_T = 0.031$ for 
$\Delta^2 (k = 0.03 {\,\rm s/km}, z=2.72) = 0.94$, and $\sigma_T=0.028$ for $\Delta^2 (k = 0.03 {\,\rm s/km}, z=2.72) = 0.58$.  The variance is similar between the different
normalizations because each normalization requires a different $A_\alpha$
in eq. (\ref{taualpha}) to match the mean transmission.  This
difference in $A_\alpha$ probably compensates for the effect
of the different normalizations on $\sigma_T$.
The predicted scatter of $\sigma_T \sim 0.030$ is consistent with the measured $\sigma_T$ of $0.03 \pm 0.01$. 

We apply the same methodology as the above to estimate $\sigma_T$
at $z \sim 6$. In Fig. \ref{sigmaT}, we show the results
for a range of different $J_{\rm HI}$'s for each of our canonical power
spectrum normalizations, 
($\Delta^2 (k = {\,\rm 0.03} {\, \rm s/km}, z=2.72) = 0.58, 0.74$ and 
$0.94$).  
\footnote{The comparison across normalizations is
done here at fixed $J_{\rm HI}$ while at $z = 5.5$ we compared the results of
different normalizations at fixed mean transmission. We find that the
dependence on normalization is larger at fixed $J_{\rm HI}$ than at fixed
mean transmission.}
Photon noise is not included in the estimates of this figure.
Even for relatively 
large $J_{\rm HI}$'s the scatter is small.  
For instance, for $J_{\rm HI} = 4.5 \times 10^{-2}$,
$\sigma_T = 1.1 \times 10^{-2}$, assuming our fiducial normalization.        
This $J_{\rm HI}$ is large in that it already gives a mean 
transmission, $\bar{f}  = 1.75 \times 10^{-2}$, in
excess of the observations.  By $J_{\rm HI} = 1.4 \times 10^{-2}$, the
scatter is only $\sigma_T = 2.8 \times 10^{-3}$ for our fiducial
normalization.  The scatter depends somewhat on normalization, as
one can see in the figure.  To measure
the scatter well would require data that are less noisy than the one
discussed here, which has photon noise of $(\sigma_n^2 / N)^{0.5} \sim 0.003$,
comparable to the predicted scatter.

On the other hand, the smallness of this scatter makes it a possibly interesting diagnostic.
As we have emphasized before, this predicted scatter ignores fluctuations
in the ionizing background. For sufficiently small $J_{\rm HI}$'s, 
the IGM should be close to the epoch of reionization, and one
would expect large sightline by sightline variations. An observed scatter
well in excess of what is predicted would be an interesting signature.

\section{Discussion}
\label{conclusions}

Our findings are summarized as follows.

\begin{figure}[htb]
\centerline{\psfig{figure=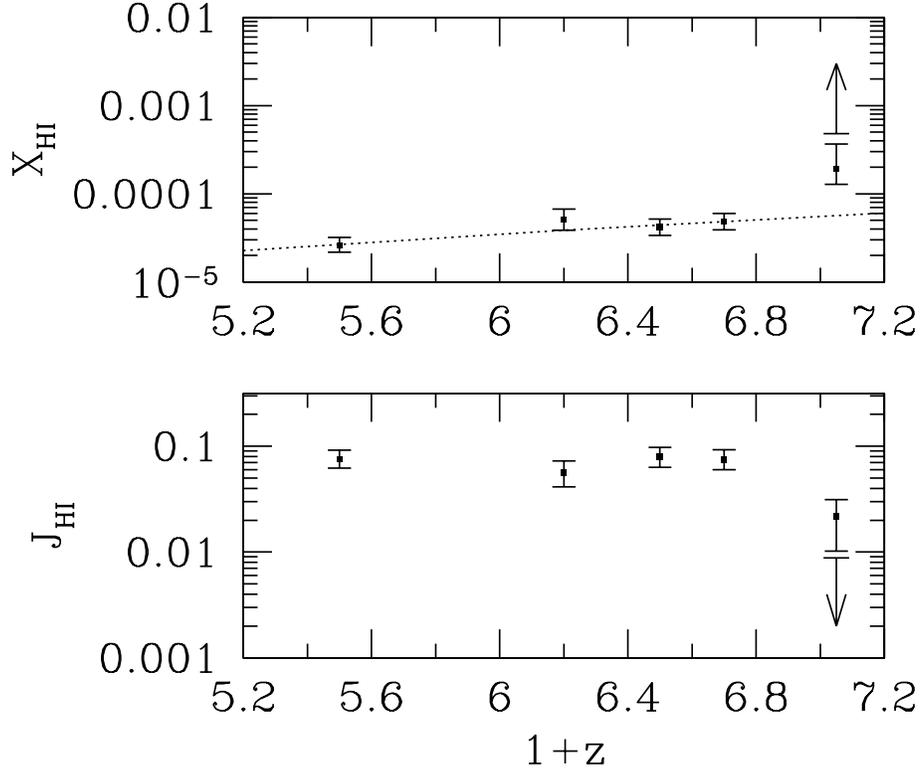,height=5.0in}}
\caption{\label{meanTalpha}
The top panel shows the neutral fraction of hydrogen
at mean density as a function of redshift implied by the measurements
of the mean transmission in the Ly$\alpha$ forest.  The point with the
error bar pointing towards a completely neutral IGM comes from matching the
mean transmission in Ly$\beta$.  The error
bars include the 1 $\sigma$ uncertainty in power spectrum normalization
and the 1 $\sigma$ error in the observed mean transmission.  The dotted line 
is offered as a guide to the eye.  It shows 
$X_{\rm HI} = 3.5 \times 10^{-5} (1+z/6)^3$.  The bottom
panel shows the corresponding evolution in the ionizing background. 
}
\end{figure}

\begin{figure}[htb]
\centerline{\psfig{figure=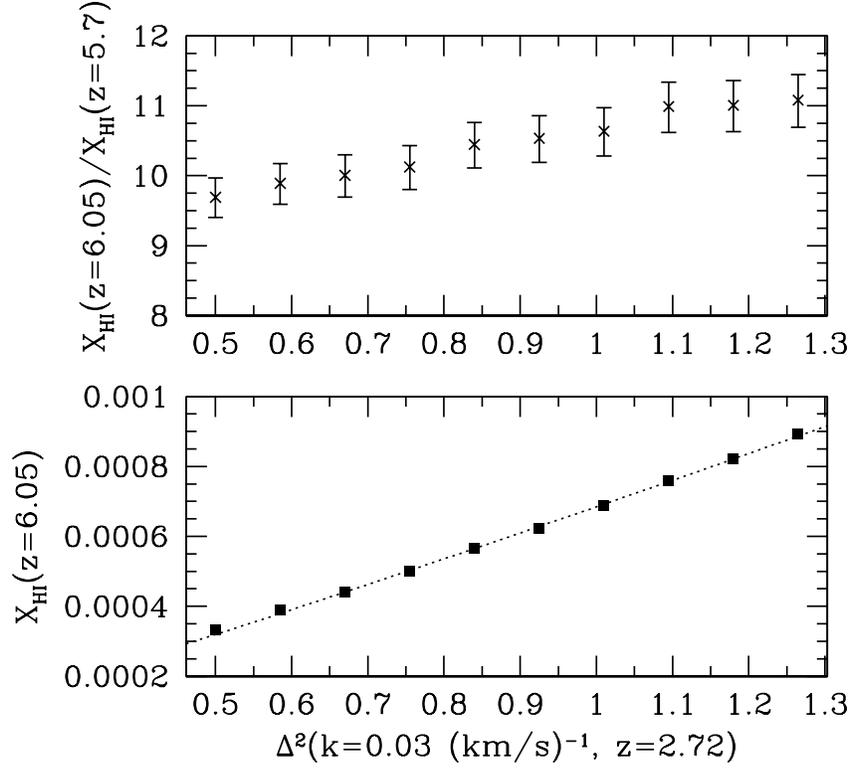,height=5.0in}}
\caption{\label{xbreak}
The upper panel shows the size of the jump in neutral fraction,
$X_{\rm HI}(z=6.05)/X_{\rm HI}(z=5.7)$, as a function of power spectrum
amplitude.  The amplitude is described by the value of
$\Delta^2 (k) \equiv 4 \pi k^3 P(k) / (2\pi)^3$ at $z = 2.72$ and
velocity scale $k = 0.03 ({\,\rm km/s})^{-1}$.  $X_{\rm HI}(z=6.05)$ corresponds
to the lower limit arising from the 1 $\sigma$ error in the mean transmission. The error bars come from
the 1 $\sigma$ uncertainty in the mean transmission at $z = 5.7$.
The lower panel shows the neutral fraction itself at $z=6.05$.  The
dotted line is $X_{\rm HI} = 5.8 \times 10^{-4} (\Delta^2 (k)/0.86)^{1.1}$,
demonstrating how the neutral fraction scales with amplitude.
}
\end{figure}

\begin{figure}[htb]
\centerline{\psfig{figure=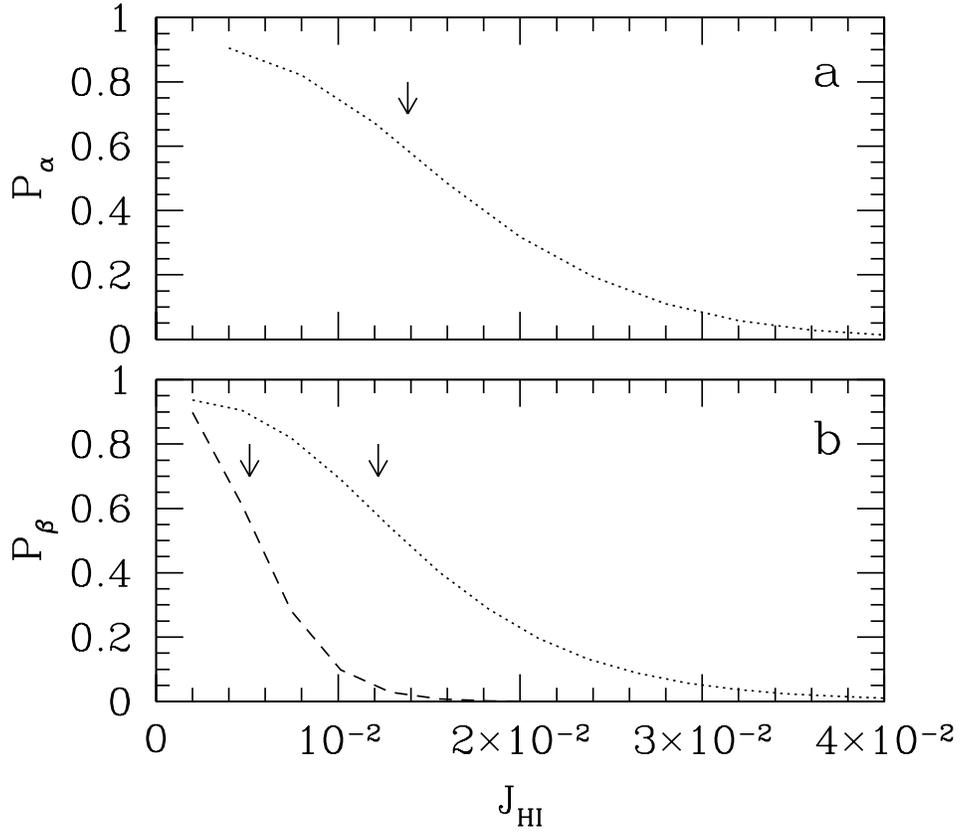,height=5.0in}}
\caption{\label{problong} The upper panel a shows the probability
$P_\alpha \equiv \int_{< F_m} P(F_1 ... F_N) dF_1 ... dF_N$ where
$F_i$ is the Ly$\alpha$ transmission at each pixel $i$ of width
$4 \AA$. Here, $N = 60$, the noise per pixel is 
$\sqrt{\langle n^2 \rangle} = 0.02$ and $F_m = 3 \sqrt{\langle n^2 \rangle}$. 
The lower panel b shows an analogous probability 
$P_\beta$ except that $F_i$ now contains both Ly$\alpha$ and
Ly$\beta$ absorption. Here, $N = 48$, $\sqrt{\langle n^2 \rangle} = 0.02$
and $0.005$ for dotted and dashed line respectively. 
The arrows indicate the corresponding 1 $\sigma$ upper limit
on $J_{\rm HI}$ for these different probability distributions.
}
\end{figure}

\begin{figure}[htb]
\centerline{\psfig{figure=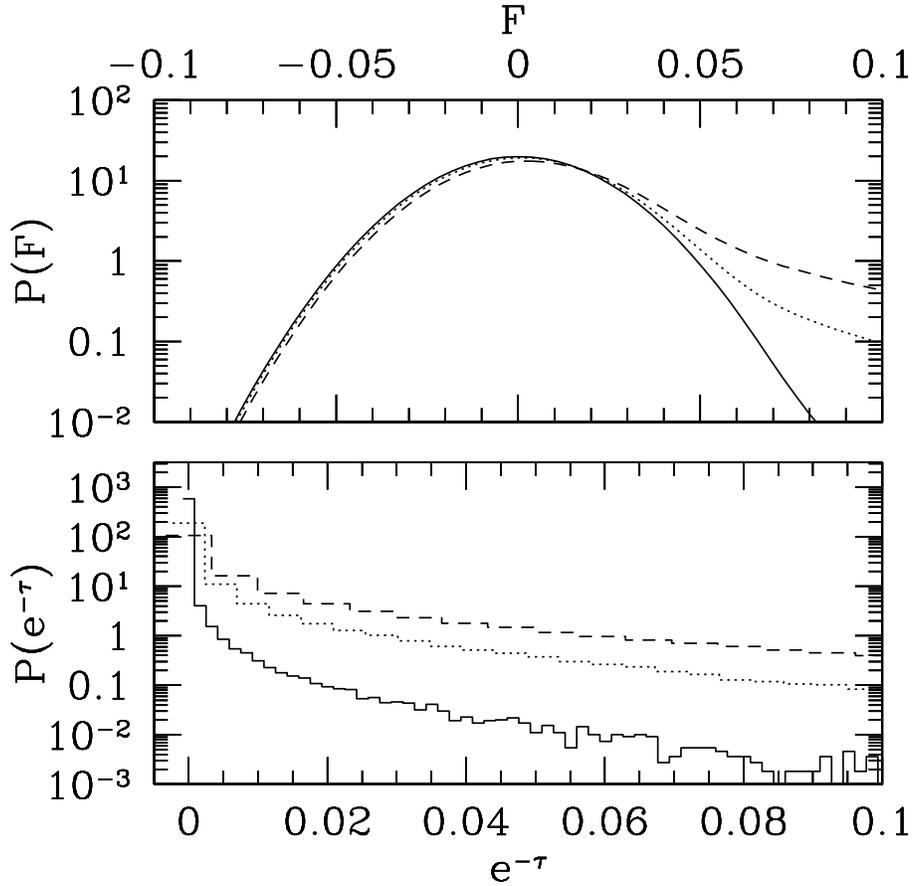,height=5.0in}}
\caption{\label{pdf} 
The lower panel shows the one-pixel ($4 \AA$) probability distribution function of 
the true noiseless transmission $e^{-\tau}$ 
(i.e. $P(e^{-\tau}) de^{-\tau}$ gives the probability) for
3 different values of $J_{\rm HI}$: 0.004 (solid), 0.012 (dotted)
and 0.028 (dashed). The upper panel shows the probability distribution
function of the noisy observed transmission $F$ for
the same three values of $J_{\rm HI}$. The negative values for $F$ occur
because of sky subtraction.
}
\end{figure}

\begin{figure}[htb]
\centerline{\psfig{figure=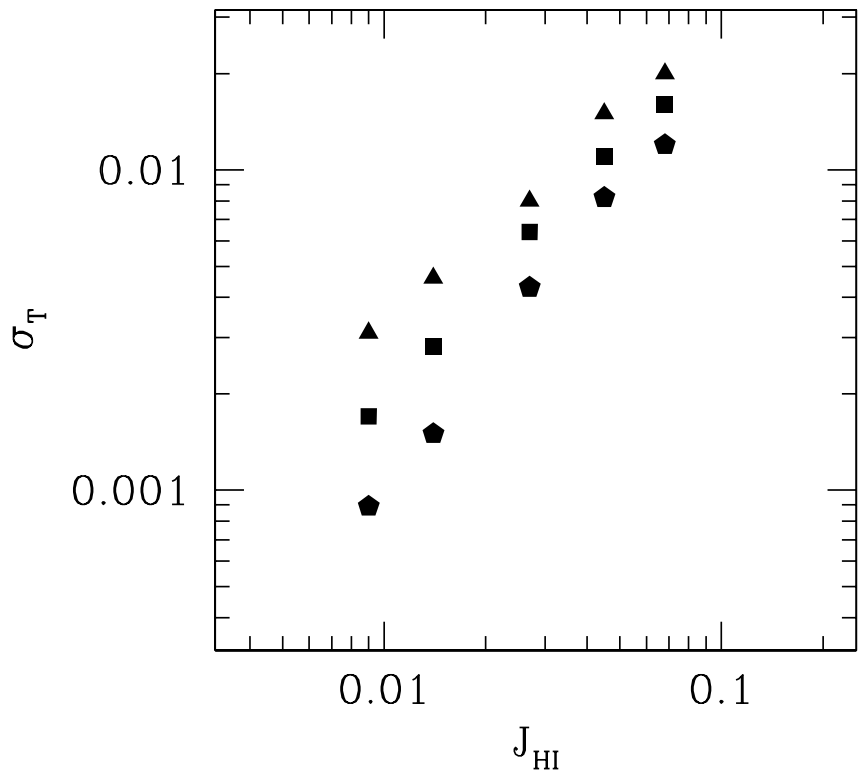,height=5.0in}}
\caption{\label{sigmaT}
A prediction of the variance of the mean transmission, $\sigma_T$,  
(see Section \ref{variance of mean t}) at $z \sim 6$, 
for several values of the ionizing
background, $J_{\rm HI}$.  The estimate ignores contributions from photon 
noise.  The triangles are for a model with power spectrum normalization
$\Delta^2 (k = 0.03 {\,\rm s/km}, z=2.72) = 0.94$, the squares 
$\Delta^2 (k = 0.03 {\,\rm s/km}, z=2.72) = 0.74$, and the pentagons
$\Delta^2 (k = 0.03 {\,\rm s/km}, z=2.72) = 0.58$.
}
\end{figure}

\begin{itemize}

\item The most stringent (1 $\sigma$) lower limit on the neutral hydrogen fraction
$X_{\rm HI}$ (eq. [\ref{XHI}]) or upper limit on the ionizing background
$J_{\rm HI}$ (eq. [\ref{Gamma}]) at $z \sim 6$ is obtained from
the observed mean Ly$\beta$ transmission: $X_{\rm HI} > 4.7 \times 10^{-4}$.
A comparison of this limit versus constraints at lower redshifts is
presented in Fig. \ref{meanTalpha}. The fact that the neutral fraction increases
by a factor of $ \sim 10$ from redshift of $5.7$ to $6$ even though it changes
by no more than a factor of about $2$ from $z=4.5$ to $z=5.7$ suggests that $z \sim 6$ might
be very close to the epoch of reionization. We emphasize that
current constraints are still consistent with a highly ionized IGM at
$z \sim 6$ -- it is the steep rise in $X_{\rm HI}$ that is suggestive of
dramatic changes around or just before that redshift.
We should also mention that the constraints on $X_{\rm HI}$ are less
subject to uncertainties in the IGM temperature compared to those on $J_{\rm HI}$
(see \S \ref{method}). 


\item The existence of a long Gunn-Peterson (Ly$\alpha$ or Ly$\beta$) 
trough at $z \sim 6$, where little or no flux
is detected, can also be used to obtain constraints on
$X_{\rm HI}$ or $J_{\rm HI}$. This we call the fluctuation method:
the fact that a long stretch of the spectrum exhibits no
large upward fluctuations in transmission provides interesting
information on the neutral fraction or ionizing background. 
The constraints obtained this way turn out
to be fairly similar to those obtained using the mean transmission. 
We estimate that a reduction in noise by a factor of 4, or
an increase in number of sightlines to 5, would result in
constraints that are 2 times stronger (\S \ref{los constraint}). 

\item We develop a method to predict the dispersion in mean transmission measured
from sightlines that are longer than the typical simulation box (eq. [\ref{sigmaTeq}]
and Fig. \ref{sigmaT}). Our predicted dispersion is consistent with that
observed at $z = 5.5$ (Becker et al. 2001). We also predict the scatter at 
redshift $z = 6$, which can be measured when more sightlines become available. 
Assuming a spatially homogeneous ionizing background, we predict 
a small scatter at $z=6$, $\sigma_T \sim$ a few $\times 10^{-3}$, neglecting
photon noise.
The dispersion provides a useful diagnostic of fluctuations in the ionizing background
-- close to the epoch of reionization, one expects large fluctuations from one line
of sight to another depending on whether it goes through regions of the IGM
where percolation of HII regions has occurred. 

\end{itemize}

There are at least three issues that will be worth exploring. 
First, with more quasars at $z \sim 6$ or higher discovered in the future, 
applying some of the ideas mentioned above would be extremely interesting,
such as the measurement of the line of sight scatter in mean transmission, or
the use of the Gunn-Peterson trough to obtain stronger constraints on the
neutral fraction. Second, as we have commented on before, fluctuations
in the ionizing background are expected to be important as we near the
epoch of reionization. 
We have not discussed it here, but a calculation
of the size of these fluctuations would be very interesting. Such a
calculation will depend both on the mean free path of the ionizing photons
as well as the spatial distribution of ionizing sources. 
The latter is probably quite uncertain, but useful estimates might be made
(e.g. Razoumov et al. 2001).
Lastly, a main source of systematic error which we have not discussed is
the continuum placement. The mean transmissions at various redshifts given by Becker et al.
are all obtained by extrapolating the continuum from the red side
of Ly$\alpha$ by assuming a power law of $\nu^{-0.5}$. 
The continuum likely fluctuates from one quasar to another, and 
therefore, it would be very useful to apply exactly the same procedure
to quasars at lower redshifts where the continuum on the blue side can be more reliably
reconstructed. This will tell us how much scatter (and possibly systematic
bias) the continuum placement procedure introduces to the measured mean 
transmission. This kind of error is especially important to quantify given the limited 
number of quasars available for high redshift measurements at the moment.

AL and LH are supported in part by the Outstanding Junior Investigator Award 
from
the DOE, and  AST-0098437 grant from the NSF. MZ is supported by NSF 
grants AST 0098606 and PHY 0116590 and by the David and Lucille 
Packard Foundation Fellowship for Science and Engineering.   
We thank Nick Gnedin for the use
of an HPM code and Rupert Croft for providing us with a revised
draft of his paper.

\newpage

\begin{table}
\begin{center}
\begin{tabular}{|cc|}\hline
z & $\langle f \rangle$ \\ \hline
$4.5$ & $0.25$ \\
$5.2$ & $0.09 \pm 0.02$ \\ 
$5.5$ & $0.097 \pm 0.002$ \\
$5.7$ & $0.070 \pm 0.003$ \\
$6.05$ & $0.004 \pm 0.003$ \\ \hline
\end{tabular}
\end{center}
\caption{\label{meanttab} A summary of the observed mean transmission.
The observation at redshift $4.5$ is from Songaila et al. (1999).  For
this observation no error bars were provided by the authors.  The observation
at $5.2$ is from Fan et al. (2000).  The other observations are from Becker et
al. (2001).  Becker et al. (2001) have two observations at $z=5.5$.  The above
mean transmission at $z=5.5$ is the average of these two observations.}
\end{table}

\newpage

\newpage

\end{document}